\documentclass[aps,prl,twocolumn,superscriptaddress]{revtex4}
\usepackage{graphics,epsfig}

\begin{document}


\title{Unusual Single-Ion Non-Fermi Liquid Behavior in Ce${}_{1-x}$La${}_x$Ni${}_9$Ge${}_4$}


\author{U. Killer}
\affiliation{CPM, Institut f\"ur Physik, Universit\"at Augsburg,
86135 Augsburg, Germany}
\author{E.-W. Scheidt}
\affiliation{CPM, Institut f\"ur Physik, Universit\"at Augsburg,
86135 Augsburg, Germany}
\author{G. Eickerling}
\affiliation{CPM, Institut f\"ur Physik, Universit\"at Augsburg,
86135 Augsburg, Germany}
\author{H. Michor}
\affiliation{Institut f\"ur Festk\"orperphysik, TU Wien, 1040 Wien,
Austria}
\author{J. Sereni}
\affiliation{ Centro Atomico Bariloche, 8400 San Carlos de
Bariloche, Argentina}
\author{Th. Pruschke}
\affiliation{Institut f\"ur Theoretische Physik, Universit\"at
G\"ottingen, 37077 G\"ottingen, Germany}
\author{S. Kehrein}
\affiliation{TP III -- EKM,
Institut f\"ur Physik, Universit\"at Augsburg, 86135 Augsburg, Germany}


\date{\today}

\begin{abstract}
We report on specific heat, magnetic susceptibility and
resistivity measurements on the compound
Ce${}_{1-x}$La${}_x$Ni${}_9$Ge${}_4$ for various concentrations
ranging from the stoichiometric system with $x=0$ to the dilute
limit $x=0.95$. Our data reveal single-ion scaling with the
Ce-concentration and the largest ever recorded value of the
electronic specific heat $\Delta c/T \approx$ 5.5\,J\,$\rm
K^{-2}mol^{-1}$ at $T=0.08$\,K for the stoichiometric compound
$x=0$ without any trace of magnetic order. While in the doped
samples $\Delta c/T$ increases logarithmically below $3$\,K down
to $50$\,mK, their magnetic susceptibility behaves Fermi liquid
like below $1$\,K. These properties make the compound
Ce${}_{1-x}$La${}_x$Ni${}_9$Ge${}_4$ a unique system on the
borderline between Fermi liquid and non-Fermi liquid physics.
\end{abstract}


\maketitle



The investigation of metals with strong correlations among the
electrons is a fundamentally important topic in modern solid state
physics \cite{Hewson}. Landau's Fermi liquid theory incorporates the effect of
electronic interactions into a renormalized electron mass $m^*$
and is the paradigm for understanding low-temperature properties
of metals. Since specific heat and magnetic susceptibility are
proportional to $m^*$, a large enhancement of $m^*$ over the free
electron mass $m_0$ leads to large values of the specific heat and
magnetic susceptibility in the heavy fermion (HF) systems
\cite{stewart84}. Measurements of low-temperature specific heat
and susceptibility can therefore reveal whether electronic
correlations renormalize the Fermi liquid parameters or lead to
 non-Fermi liquid (nFL) behavior \cite{stewart01,Scho99}.
The breakdown of Fermi liquid theory and the borderline between
these regimes continues to attract much interest.

\begin{figure}
\centerline{\includegraphics[width=6cm,clip]{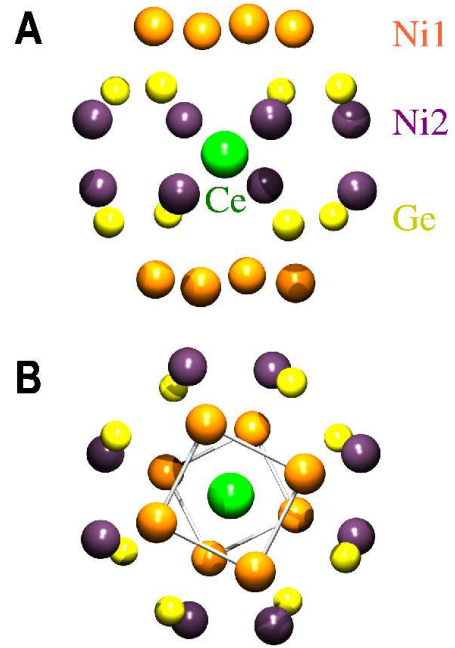}}
\caption{A simplified representation of the threefold tetragonal
antiprismatic environment of Ce: A) view along the
crystallographic a-axis; B) view along the c-axis. In B the
tetragonal faces of the antiprism formed by the eight Ni1 located
at the 16k sites are indicated by squares which are rotated
against each other by approximately 35$^{\circ}$. Notice that the
Ni2 and Ge reside at places with the same site symmetry (16l)
\cite{michor_inpress}.}
\label{fig1}                          
\end{figure}

In this work we report on specific heat and magnetic
susceptibility measurements on the HF-compound CeNi$_9$Ge$_4$
\cite{michor_inpress} that show pronounced nFL-behavior of the
specific heat over one-and-a-half decade of temperature. In fact,
this Ce-f-electron lattice system turns out to have the largest
ever recorded value of the electronic specific heat at low
temperature: $\Delta c/T \approx$ 5.5\,J\,$\rm K^{-2}mol^{-1}$ at
0.08\,K, without showing any trace of magnetic order (Only in the
magnetic YbBiPt compound a higher$\Delta c/T \approx$ 8\,J\,$\rm
K^{-2}mol^{-1}$ is observed \cite{Fisk}). This Sommerfeld
coefficient exceeds considerably the values observed in other
non-magnetic (e.g. CeAl$_3$, CeCu$_6$ or CeCuIn$_2$) or magnetic
compounds (e.g. CeAl$_2$, CePb$_3$ or CeAgIn$_2$) with $c/T \leq$
1.6\,J\,$\rm K^{-2}mol^{-1}$ \cite{stewart84,Sereni91}, and even
the $c/T$ value of Ce-systems tuned to their critical points (e.g.
CePd$_3$B, with $c/T_{T \rightarrow 0} \leq$ 3.38\,J\,$\rm
K^{-2}mol^{-1}$ \cite{Sereni91}, CeCu$_x$Al$_y$,
CeCu$_{6-x}$Au$_x$, CePd$_{1-x}$Ni$_x$ or Ce$_7$Ni$_3$ under
pressure \cite{Sereni01}).

\begin{figure}
\centerline{\includegraphics[width=8cm,clip]{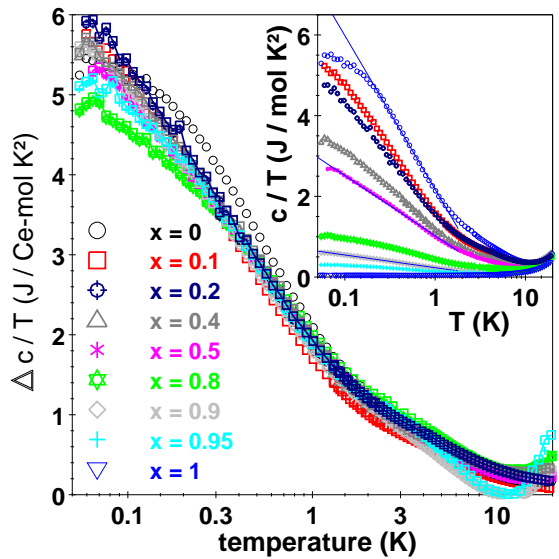}}
\caption{The electronic contribution to the specific heat
$\Delta$c
 of Ce${}_{1-x}$La${}_x$Ni${}_9$Ge${}_4$ divided by temperature
and normalized per Ce-mol. The inset shows the specific heat
divided by temperature of Ce${}_{1-x}$La${}_x$Ni${}_9$Ge${}_4$.
The three solid lines represent fits to the data with $c/T =
\gamma_{0} - a~\ln~T$ below 1.5\,K.}
\label{fig2}                          
\end{figure}

Remarkably, our experiments reveal that the nFL-like logarithmic
increase of the electronic specific heat $\Delta c/T$ in
CeNi$_9$Ge$_4$ (0.3K $<$ T $<$ 1.5K) is proportional to the
Ce-concentration in La-substituted samples
Ce${}_{1-x}$La${}_x$Ni${}_9$Ge${}_4$, which suggests that a
single-ion effect due to the Ce-ions is responsible for the
exceptional low-temperature properties of the doped compounds.
While these observations preclude collective effects like the
vicinity to a quantum phase transition as the cause of this
nFL-behavior, the low-temperature behavior of
Ce${}_{1-x}$La${}_x$Ni${}_9$Ge${}_4$ also poses a theoretical
challenge in the framework of single-ion models: it is difficult
to theoretically reconcile the observed nFL-behavior in the
electronic specific heat with a Fermi liquid-like behavior of the
magnetic susceptibility. The exceptional properties of the
compound  Ce${}_{1-x}$La${}_x$Ni${}_9$Ge${}_4$ are therefore not
only remarkable from a materials science point of view, but also
make it an interesting testing ground for theories of strongly
correlated electronic systems.

All polycrystalline samples Ce${}_{1-x}$La${}_x$Ni${}_9$Ge${}_4$
presented in this work were prepared by arc-melting of pure
elements (Ce, 4N; La, 4N; Ni, 4N7; Ge, 6N) under a highly purified
argon atmosphere. To obtain the highest possible homogeneity, the
samples were flipped over four times and remelted. Subsequently
the samples were annealed in an evacuated quartz glass tube for
seven days at 1000$^{\circ}$C. Powdered samples were investigated
by standard X-ray techniques using Cu$K_{\alpha}$ radiation.
CeNi${}_9$Ge${}_4$ crystallizes in a tetragonal structure with
space group I4/mcm and lattice parameters  a = 7.9701(1)${\AA}$
and c = 11.7842(3)${\AA}$ \cite{michor_inpress}. The coordination
of the Ce atoms is depicted in Fig.~\ref{fig1}. Each Ce atom has
the same threefold tetragonal antiprismatic  environment formed by
sixteen Ni and eight Ge neighbors. Replacement of the Ce-atoms by
La leads to a slight volume expansion of about 0.6\%. Both lattice
parameters follow Vegard's law.

The dc-susceptibility measurements were performed with a
commercial SQUID magnetometer for temperatures 1.8\,K\,\,$< T
<$\,\,400\,K, and were completed in the low temperature region
(0.03\,K\,\,$< T <$\,\,2.5\,K) by ac-susceptibility measurements
in a $^3$He-$^4$He-dilution refrigerator. The specific heat
experiments were conducted in noncommercial setups using a
standard relaxation method \cite{Bachmann} in a conventional
$^{4}$He-cryostat (1.8\,K\,\,$< T <$\,\,70\,K), and in a
$^3$He-$^4$He-dilution refrigerator at low temperatures
(0.05\,K\,\,$< T <$\,\,2.5\,K).

The specific heat divided by temperature  of
Ce${}_{1-x}$La${}_x$Ni${}_9$Ge${}_4$(with $x$ ranging between 0
and 1) is displayed in the inset of Fig.~\ref{fig2} in the
temperature range 0.05\,K\,\,$< T <$\,\,20\,K. For all
La-substituted samples a nearly logarithmic increase of  $ c/T$
below 1.5\,K is observed which is characteristic of nFL-physics.
Only the non-diluted compound CeNi${}_9$Ge${}_4$ deviates
noticeably from this logarithmic behavior below 300\,mK. Notice
that this poly-crystalline sample reaches nearly the same value of
5.5\,J\,$\rm K^{-2}mol^{-1}$ for the Sommerfeld coefficient at $T
= 0.08$\,K as observed for single crystals \cite{michor_inpress}.

In order to extract the electronic contribution to the specific
heat, we measured the non f-electron system LaNi${}_9$Ge${}_4$
from which we derived the lattice vibration contribution. The
phonon contribution can be well parameterized using a Debye term
and two Einstein modes. In this calculation we fixed the number of
internal degrees of freedom to 3x14=42, according to the 14 atoms
in the unit cell. The  Debye temperature $\Theta_{D} = 123$\,K and
two Einstein temperatures  $\Theta_{E} = 187$\,K and 440\,K were
calculated with a weight distribution 3:24:15, respectively.

\begin{figure}
\centerline{\includegraphics[width=8cm]{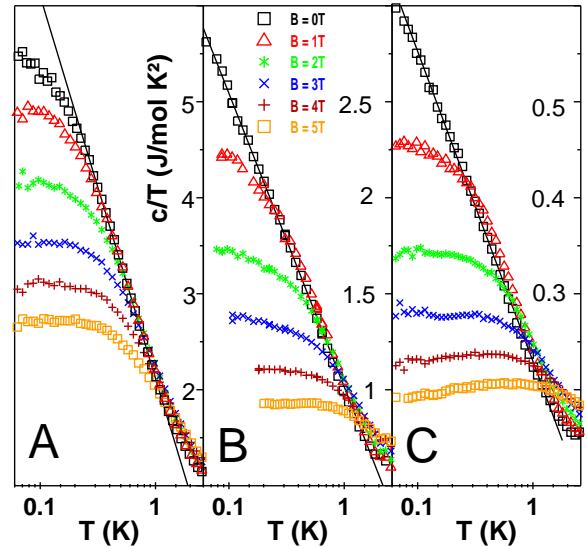}}
\caption{The electronic specific heat divided by temperature in
various magnetic fields of three particular samples A)
CeNi${}_9$Ge${}_4$ ; B) Ce$_{0.5}$La$_{0.5}$Ni${}_9$Ge${}_4$ and
C) Ce$_{0.1}$La$_{0.9}$Ni${}_9$Ge${}_4$. The solid lines are
logarithmic  fits to the zero-field data below 1.5\,K.}
\label{fig3}                          
\end{figure}

The electronic contribution to the specific heat $\Delta c/T$ was
then obtained by subtracting this phonon contribution from the
measured $c/T$. The resulting curves $(\Delta c/T)/(1-x)$
normalized per Ce-concentration are displayed in Fig.~\ref{fig2}.
The record value of the electronic Sommerfeld coefficient of
nearly 5.5\,J\,$\rm K^{-2}mol^{-1}$ is found to be almost
Ce-concentration independent. Even the slightly lower values below
0.2\,K for $x = 0.8$ and 0.95 are consistent with this value in
spite of the large error introduced by dividing the experimental
data by the small number $1-x$. The curves in Fig.~\ref{fig2}
provide compelling evidence for scaling behavior with the
Ce-concentration; only the non-dilute compound with $x=0$ deviates
systematically (probably due to collective interactions). This
strongly suggests that the low-temperature physics in
Ce$_{1-x}$La$_{x}$Ni$_{9}$Ge$_{4}$ is governed by single-ion
behavior due to the Ce-ions.

The logarithmic increase of the Sommerfeld coefficient flattens
off and crosses over into Fermi liquid-like behavior in an
external magnetic field: such measurements of the specific heat
for the compounds with $x =0$, 0.5, and 0.9 in various applied
magnetic fields are depicted in Fig.~\ref{fig3}. This confirms a
single-ion Kondo-like scenario with strong electronic correlations
due to the magnetic moment of the Ce${}^{3+}$-ions.

\begin{figure}
\centerline{\includegraphics[width=8cm]{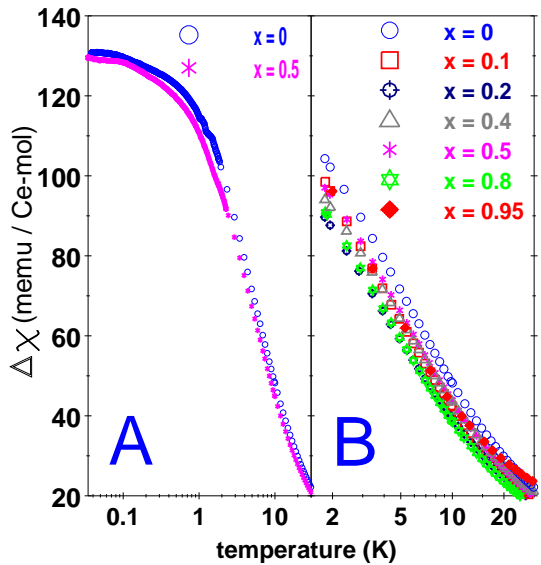}}
\caption{The local magnetic susceptibility $\Delta\chi$ of
Ce${}_{1-x}$La${}_x$Ni${}_9$Ge${}_4$ normalized per Ce-mol: A)
Low-temp\-erature data for samples x=0 and x=0.5 down to $0.03 K$
obtained by normalizing the ac-susceptibility data (magnetic field
$<$\,0.3mT) to the dc-susceptibility data between $1.8 K$ and $2.5
K$. B) dc-susceptibility data in 0.5T for various concentrations
x. }
\label{fig4}                          
\end{figure}

\begin{figure}
\centerline{\includegraphics[width=8cm]{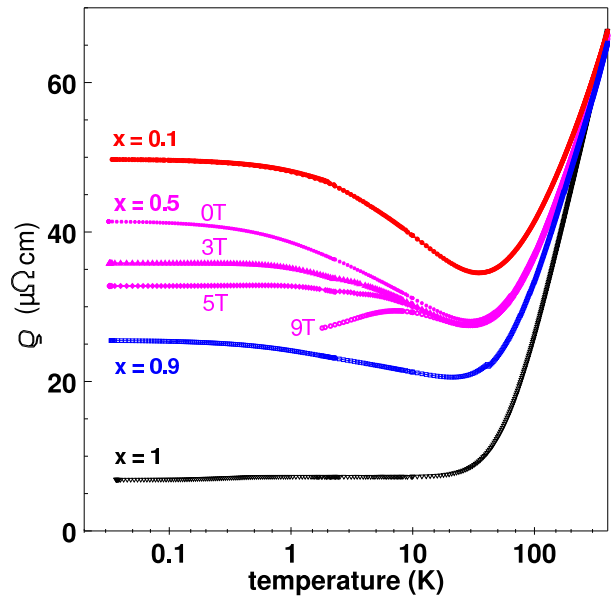}}
\caption{The electrical resistivity $\rho(T)$ of various samples
normalized to that of LaNi${}_9$Ge${}_4$ at 300\,K (Vander-Pauw
method). The data for $x=0.5$ shows the reduction of the residual
resistivity with an external magnetic field.}
\label{fig5}                          
\end{figure}

Measurements of the magnetic susceptibility for various values of
the Ce-concentration are presented in Fig.~\ref{fig4}B. Similar to
the specific heat analysis, we have subtracted the magnetic
susceptibility of LaNi$_{9}$Ge$_{4}$; the resulting curves again
exhibit single-ion scaling with respect to the Ce-concentration.
Like the electronic contribution to the specific heat, the local
magnetic susceptibility presented in Fig.~\ref{fig4} also shows a
large enhancement over the free electron value with a magnitude
comparable to the specific heat data. In particular, $\Delta\chi$
also increases logarithmically for temperatures 2\,K\,\,$< T
<$\,\,10\,K. However, in marked contrast to the specific heat the
local susceptibility $\Delta\chi$ flattens off and becomes
constant below 1\,K (Fig.~\ref{fig4}A), while $\Delta c/T$
continues to grow for at least another one-and-a-half decade down
to 50\,mK (compare e.g.\ Fig.~\ref{fig3}C for B=0). This
discrepancy between the nFL-like behavior of the Sommerfeld
coefficient and the Fermi liquid-like behavior of the magnetic
susceptibility is one of the key observations of our work and
leads to a strongly temperature-dependent Wilson ratio below $1K$.

Resistivity measurements were performed over four decades in
temperature below 300\,K for selected samples (Fig.~\ref{fig5}).
LaNi${}_9$Ge${}_4$ exhibits normal metallic behavior with a
residual resistivity of 7~$\mu\Omega$cm. With increasing
Ce-concentration a Kondo-like resistivity minimum is formed out,
leading to a very high residual resistivity at 30\,mK (e.g.\
50~$\mu\Omega$cm for $x = 0.1$). This residual resistivity can be
reduced in an external magnetic field (see Fig.~\ref{fig5}), which
indicates Kondoesque magnetic correlations. The electrical
resistivity of all the La-substituted samples exhibits single-ion
local non-Fermi liquid-like behavior with $\rho(T)-\rho(0) \propto
T^{c}$, c = 0.8$\pm$0.2 below 2\,K.

The above experimental observations effectively rule out two of
the three theoretical routes that are typically invoked for
describing nFL-like behavior: vicinity to a quantum phase
transition \cite{Andraka} and/or a disorder distribution of Kondo
temperatures \cite{disorder}. i)~A quantum phase transition
scenario with collective magnetic excitations does not lead to the
observed single-ion scaling in the dilute limit, rather the
collective excitations should disappear and the nFL-behavior
vanish in the dilute limit $x\rightarrow 1$. ii)~A Kondo disorder
description would imply that the disorder distribution in our
compound is essentially unaffected by the dilution, which is
physically unrealistic. Also, Kondo disorder models lead to
logarithmic behavior in both the Sommerfeld coefficient and the
magnetic susceptibility in the same temperature range, which is
different from the observations in
Ce${}_{1-x}$La${}_x$Ni${}_9$Ge${}_4$. We therefore propose a local
single-ion nFL-scenario, which makes
Ce${}_{1-x}$La${}_x$Ni${}_9$Ge${}_4$ with its large value of the
Sommerfeld coefficient a unique testing ground for unconventional
Kondo models (e.g.\ 2-channel Kondo models)
\cite{footnote1,cox98}.

In order to better understand the local physics, we have
calculated the entropy based on our specific heat data including
higher temperatures. These calculations show that R~ln2 per Ce-mol
is already reached at about 3\,K, which signals the participation
of more than two degrees of freedom per Ce atom relevant for the
thermodynamic behavior below $T = 15$\,K. A possible origin for
this is a ground state quartet of Ce${}^{3+}$ that is split into
two doublets by the distorted tetragonal antiprismatic crystal
field around each Ce atom formed by the Ge and Ni atoms (see
Fig.~\ref{fig1}). This can lead to an interplay between the Kondo
effect and crystal field splitting on the same energy scale.

This model has been investigated theoretically by Desgranges and
Rasul \cite{Desgranges}. While the Sommerfeld coefficient always
becomes finite (Fermi liquid like) for $T\rightarrow 0$ in this
model, one can imagine that the Kondo temperature is too small in
the Ce$_{1-x}$La$_{x}$Ni${}_9$Ge${}_4$ system to observe the
flattening off in the experimental data. However, the flattening
off of the magnetic susceptibility curve below $1K$
(Fig.~\ref{fig4}A) {\em in contrast} with the increasing $\Delta
c/T$-values (see e.g. Fig.~\ref{fig3}C for B=0) is very puzzling
and difficult to reconcile with a local Fermi liquid picture:
theoretically this behavior can be ruled out for large crystal
fields $\Delta_{CF}\gg T_K$ where the low-energy behavior is
governed by local Fermi liquid physics due to an effective
conventional spin-1/2 Kondo model. For vanishing crystal field we
have performed a 4-band numerical renormalization group
calculation \cite{Wilson} and found similarly that the
Desgranges-Rasul model is incompatible with the observed behavior
of $\gamma$ and $\chi$. Unfortunately, no reliable theoretical
data for the magnetic susceptibility is available for general
crystal field splittings between these two limiting cases to date.

We conclude that while single-ion correlation models are, in
general, theoretically well-understood, the exceptional behavior
of the Ce${}_{1-x}$La${}_x$Ni${}_9$Ge${}_4$ system cannot easily
be explained by any of them. In particular, the non-Fermi liquid
like behavior of the specific heat in contrast with the Fermi
liquid like behavior of the magnetic susceptibility below 1\,K
poses a challenge to theory and seems to rule out standard local
Fermi liquid descriptions. This suggests that the
Ce${}_{1-x}$La${}_x$Ni${}_9$Ge${}_4$ system is in a novel
non-Fermi liquid state that exhibits both non-Fermi liquid and
Fermi liquid-like properties. A more complex theoretical model,
which would incorporate local non-Fermi liquid physics like the
two-channel Kondo model \cite{cox98} plus low-lying crystal field
splittings, might be necessary to describe these remarkable
low-temperature properties. Of particular interest are also the
additional collective effects below 300\,mK in the non-dilute
compound ($x=0$). Further investigations of
Ce${}_{1-x}$La${}_x$Ni${}_9$Ge${}_4$ with its record value of the
specific heat could be a key to gaining a better understanding of
this borderline between Fermi liquid and non-Fermi liquid physics.

We acknowledge valuable discussions with W.~Scherer, E.~Bauer,
G.~Hilscher and P.~Rogl. This work was supported by the Austrian
FWF~15066 and through SFB~484 of the Deutsche
Forschungsgemeinschaft (DFG). SK acknowledges support through a
Heisenberg fellowship of the DFG.

\end{document}